\documentclass[12pt]{iopart}

\expandafter\let\csname equation*\endcsname\relax
\expandafter\let\csname endequation*\endcsname\relax 

\usepackage[all,2cell]{xy}
\usepackage{color}
\usepackage[latin1]{inputenc}
\usepackage{epsfig}
\usepackage{transparent}
\usepackage{nicefrac}
\usepackage{amsmath}
\usepackage{amssymb}
\usepackage{chemarrow}

\newcommand{\be}{\begin{equation}}
\newcommand{\ee}{\end{equation}}
\newcommand{\bea}{\begin{eqnarray}}
\newcommand{\eea}{\end{eqnarray}}
\newcommand{\bes}{\begin{subequations}\bea}
\newcommand{\ees}{\eea\end{subequations}}
\newcommand{\ba}{\begin{array}}
\newcommand{\ea}{\end{array}}
\newcommand{\bra}[1]{\langle \, #1 \,|}
\newcommand{\ket}[1]{|\,#1\,\rangle}
\newcommand{\braket}[2]{\langle \,#1\,|\,#2\,\rangle}
\newcommand{\bo}[1] {\boldsymbol{#1}}

\newcommand{\X}{\boldsymbol{\mathrm{X}}}
\newcommand{\x}{\mathrm{X}}
\newcommand{\y}{\mathrm{Y}}
\newcommand{\TO}{\ar@<.2ex>@{-^{>}}}
\newcommand{\FROM}{\ar@<-.2ex>@{_{<}-}}

\begin{document}
 
\title{Transient fluctuation theorems for the currents \\ and initial equilibrium ensembles}

\author{Matteo Polettini and Massimiliano Esposito}

\address{Complex Systems and Statistical Mechanics, University of Luxembourg, Campus
Limpertsberg, 162a avenue de la Fa\"iencerie, L-1511 Luxembourg (G. D. Luxembourg)}
\ead{matteo.polettini@uni.lu}

\begin{abstract}
We prove a transient fluctuation theorem for the currents for continuous-time Markov jump processes with stationary rates, generalizing an asymptotic result by Andrieux and Gaspard [J. Stat. Phys. \textbf{127}, 107 (2007)] to finite times. The result is based on a graph-theoretical decomposition in cycle currents and an additional set of tidal currents that characterize the transient relaxation regime. The tidal term can then be removed by a preferred choice of a suitable initial equilibrium ensemble, a result that provides the general theory for the fluctuation theorem without ensemble quantities recently addressed in [Phys. Rev. E {\bf 89}, 052119 (2014)]. As an example we study the reaction network of a simple stochastic chemical engine, and finally we digress on general properties of fluctuation relations for more complex chemical reaction networks.
\end{abstract}


% 05.70.Ln Nonequilibrium Thermodynamics
% 02.50.Ga Markov processes
% 82.29.-w,82.29.-s Chemical kinetics, chemical thermodynamics

% \tableofcontents

\newpage

\section{Introduction}

Fluctuation theorems (FT's in the following) have dominated the last twenty years of research in nonequilibrium statistical mechanics. Proceeding from the landmark formulation by Bochkov and Kuzovlev \cite{bochkov}, a host of variations on the theme have been elaborated depending on the theoretical setup, the observables of interest and the time specifics. This paper inscribes in the line of inquiry of FT's for stochastic dynamics \cite{kurchan,maes,lebowitz}, with special regard to the observables related to the cycle decomposition of Markov processes \cite{kalpa}.

The relevance of cycle currents and their conjugate affinities to nonequilibrium thermodynamics was investigated by Hill and Schnakenberg \cite{hill,schnak}. The intuitive picture is that a cycling process performed by a system is capable of transducing and transforming energy across the environment. As an example, the Otto cycle in the stationary performance of a car engine transforms the fuel's chemical energy into the vehicle's kinetic energy. Hence, a full characterization of the cycle structure of the system allows for the characterization of the thermodynamic behavior of nonequilibrium steady states, e.g. as regards their insurgence from a minimum entropy production principle \cite{polettini1}.
In this setting, Andrieux and Gaspard have derived an asymptotic FT for the now-called Schnakenberg cycle currents \cite{andrieux} and applied it to chemical reactions \cite{andrieux2}. Further insights on FT's and large deviations for cycle currents can be found in Refs. \cite{artur,faggionato,bertini}.
 
Under the assumption of local detailed balance \cite{espositoldb} for quantum systems coupled with several heat and particle reservoirs, upon which cycle currents acquire a simple physical interpretation, recently Bulnes-Cuetara et al. \cite{cuetara} have shown that a fluctuation relation for the currents also holds at finite times, provided that the processes are sampled from one specific initial equilibrium ensemble. We also refer to Ref.\,\cite{agmt} for some earlier results, Ref.\,\cite{park} for an analysis of heat vs. work FT's, Ref.\,\cite{fogedby} for further elaboration and Ref.\,\cite{campisi2} for the derivation of a similar result in a deterministic setting. 

In this paper we provide the general theory underlying transient FT's for time-homogeneous Markov jump processes. In particular, we generalize the result of Andrieux and Gaspard by including in the description certain tidal currents that complement the cycle currents. The result is based on an algebraic graph-theoretical analysis investigated by one of the authors in Ref.\,\cite{polettini2}. We can then generalize the initial-ensemble result, extending it to time-homogeneous Markov processes on graphs without the requirement of local detailed balance. As an example, we analyze a simple chemical reaction network.

The paper is structured as follows. In  Sec.\,\ref{note} we anticipate the forms taken by the various fluctuation relations. In Sec.\,\ref{CN1} we initialize the example of a chemical reaction network. In Sec.\,\ref{tools} we provide preliminary results from graph theory, and in Sec.\,\ref{results} we give the general results from direct manipulations of the probability density of Markov jump processes, while for completeness in \ref{appendix} the same results are derived in the Feynman-Kac formalism for the moment generating function. In Sec.\,\ref{CN2} we look back at the example under a new light, before coming to conclusions. % in Sec.\,\ref{conclusions}.

\section{\label{note}A recap on fluctuation relations}

Before moving to the full treatment, it is useful to make the statements in the introduction slightly more precise. The simplest fluctuation relation takes the form
\bea
\frac{P(\Sigma_t)}{P(-\Sigma_t)} \asymp e^{\Sigma_t}. \label{eq:psigma}
\eea
Here, $\Sigma_t$ is the value taken by a stochastic variable $\Sigma(t)$ called the reservoir entropy production of a process (sometimes denoted $\Delta _r S$, $- \Delta_e S$ etc.), which accounts for the flux of entropy towards the environment. In our setting, the entropy production is a stochastic process with probability $\mathrm{Prob}\{\Sigma(t) \in [\Sigma_t,\Sigma_t+d\Sigma_t]\} = P(\Sigma_t)d\Sigma_t$, and $\asymp$ denotes the long time limit (in the following we will not distinguish between probabilities an probability densities). Then Eq.\,(\ref{eq:psigma}) states that at sufficiently large times the probability of measuring a positive entropy production is exponentially favored with respect to the probability of measuring a negative entropy production. Since the entropy production is odd under time reversal, the fluctuation relation provides a formulation of the second law of thermodynamics and a characterization of the arrow of time.

To the entropy production of a system several mechanisms may contribute. Then, the fluctuation relation can be specialized as follows
\bea
\frac{P(\bo{J}_t)}{P(-\bo{J}_t)} \asymp e^{\bo{F}\cdot \bo{J}_t},
\eea 
where $\bo{J}_t$ are the values taken by some physical observables that (almost surely) grow linearly in time, e.g. time-integrated heat fluxes, charge or matter currents, or any thermodynamic flux. The quantities $\bo{F}$ are non-fluctuating intensive variables conjugate to the $\bo{J}_t$. If one adopts an abstract characterization of thermodynamic processes as generic Markov processes on a discrete state space, then $\bo{J}_t$ count the net number of times the process has performed certain elementary cyclic paths.

Asymptotic relations can be extended to finite times by conditioning both the forward and the backward processes to some fixed initial state \cite{seifert},
\bea
\frac{P(\bo{J}_t | x_0)}{P(-\bo{J}_t | x_t)} = e^{\bo{F}\cdot \bo{J}_t + \Phi(x_0) - \Phi(x_t)}
\eea 
where $\Phi$ is a suitable state function. Unfortunately, from an experimental viewpoint conditioning a process to one exact initial state is problematic. However, notice that if one could sample both the forward and the backward processes with probability $e^{-\Phi}/Z$ ($Z$ the normalization factor) one obtains an exact FT for the currents valid at all times
\bea
\frac{P(\bo{J}_t)}{P(-\bo{J}_t)} = e^{\bo{F}\cdot \bo{J}_t}, \label{eq:Pbo}
\eea
where we marginalized out $ x_0,x_t$. Yet, again, preparing the system in a given ensemble $e^{-\Phi}/Z$ might also be awkward, unless it is of a very special kind. Indeed, for certain classes of systems it has been found that this ensemble is the equilibrium ensemble of the system where all forces producing cycles currents are momentarily disconnected. Physically, this corresponds to the situation where first one prepares the system by letting it relax to equilibrium, and then all of a sudden connects the external forces.

\section{\label{CN1}Example: network of chemical reactions}

In this section we consider a simple reaction network. We derive a meaningful expression for the total entropy produced after an arbitrary sequence of reactions, writing it in terms of macroscopic physical currents of certain external species called {\it chemostats}, and in terms of an equilibrium initial ensemble. The reader eager to learn the full theory might want to skip this section. For sake of simplicity we set $k_B T = 1$.

Let $\x_1$ and $\x_2$ be two chemical species of observational interest that partake to three reversible chemical reactions, one that produces or consumes $\x_1$, one that produces or consumes $\x_2$, and one that converts $\x_1$ into $\x_2$ and vice versa:
\bea
\phantom{\x_1 ~+} \y_1 & ~~\autorightleftharpoons{{\scriptsize +1}}{{\scriptsize $-$1}} ~~ &  \x_1 \nonumber \\
\phantom{\x_1 ~+} \x_2   & ~~\autorightleftharpoons{{\scriptsize +2}}{{\scriptsize $-$2}} ~~ &  \y_2 \label{eq:CN} \\
\x_1 + \y_3 & ~~\autorightleftharpoons{{\scriptsize +3}}{{\scriptsize $-$3}} ~~ & \x_2 + \y_4. \nonumber
\eea
Here $\y_1,\ldots,\y_4$ are (assemblies of) chemostats, that is, substrate species that are independently administered by the environment and whose concentrations do not vary in time. A complete treatment of the thermodynamics of chemostatted networks has been provided by the authors in Ref.\,\cite{polespo}. This reaction scheme is a simple model of a molecular engine, where reactions 1 and 2 provide the working substances $\x_1$ and $\x_2$, and reaction 3 performs chemical work by transforming molecules of $\y_3$ into molecules of $\y_4$, while completing a thermodynamic cycle within the system. The observable of interest is the rate $J_3$ at which this latter reaction proceeds. The reaction network can be represented by a graph whose edges are the {\it complexes} of the species of observational interest, as follows
\bea
\ba{c}\xymatrix{
\x_1 \FROM[rr]\TO^{3}[rr]  \FROM[dr]\TO_{1}[dr] & & \x_2 \FROM[dl]\TO^{2}[dl]  \\
& \emptyset &}\ea. \label{eq:cn}
\eea

Under several assumptions (Boltzmann's Stosszahlansatz, well-stirred solution etc.), the number of variable molecules undergoes a continuous-time Markov jump process satisfying the random-time change equation
\bea
\X(t) = \X(0) + \sum_{r = \pm 1}^{\pm 3} J_r(t) \, \bo{\nu}_r,
\eea
where we collected the two variable species in a vector $\X$, and $\bo{\nu}_r$ is the vector of stoichiometric coefficients of the $r$-th reaction,
\bea
\bo{\nu}_{\pm 1} = \pm \left(\ba{c}+1 \\ 0 \ea \right), \quad \bo{\nu}_{\pm 2} = \pm \left(\ba{c}0 \\ -1 \ea \right), \quad \bo{\nu}_{\pm 3} = \pm \left(\ba{c}-1 \\ +1 \ea \right).
\eea
Each time a reaction proceeds the populations increase by an amount $\bo{\nu}_r$. Hence, the state space where this random process takes place is the lattice (that we call the {\it chemical lattice}) generated by the three vectors $\bo{\nu}_{+1}, \bo{\nu}_{+2}, \bo{\nu}_{+3}$, limited to the sector of positive populations, as depicted in Fig.\,\ref{fig:CN}(a). Notice that the generating vectors are not independent, as
\bea
\bo{\nu}_{+1} +\bo{\nu}_{+2}+\bo{\nu}_{+3} = 0. \label{eq:notind}
\eea
The quantity $J_r(t)$, counting the number of times reaction $r$ occurs up to time $t$, is distributed with a unit-rate Poisson distribution \cite{kurtz} according to 
\bea
J_r(t) \sim \mathrm{Pois} \left( \int_0^t w_{\X(s)+\bo{\nu}_r,\X(s)} \, ds \right). \label{eq:pois}
\eea 
The $\{w_{\X+\bo{\nu}_r,\X}\}_{\X}$'s are the rates at which reaction $r$ proceeds. By the law of mass-action these rates are proportional to the products of the abundances of the reactants,
\begin{align}
w_{\X+\bo{\nu}_1,\X} & = \y_1, &
w_{\X+\bo{\nu}_2,\X} & = \x_2, &
w_{\X+\bo{\nu}_3,\X} & = \y_3 \x_1, \nonumber \\
w_{\X-\bo{\nu}_1,\X} & = \x_1, &
w_{\X-\bo{\nu}_2,\X} & = \y_2, &
w_{\X-\bo{\nu}_3,\X} & = \y_4 \x_2,
\end{align}
where for sake of simplicity we set all proportionality constants to unity. 
	\begin{figure}[t]
	\centering
	\def\svgwidth{300pt}
	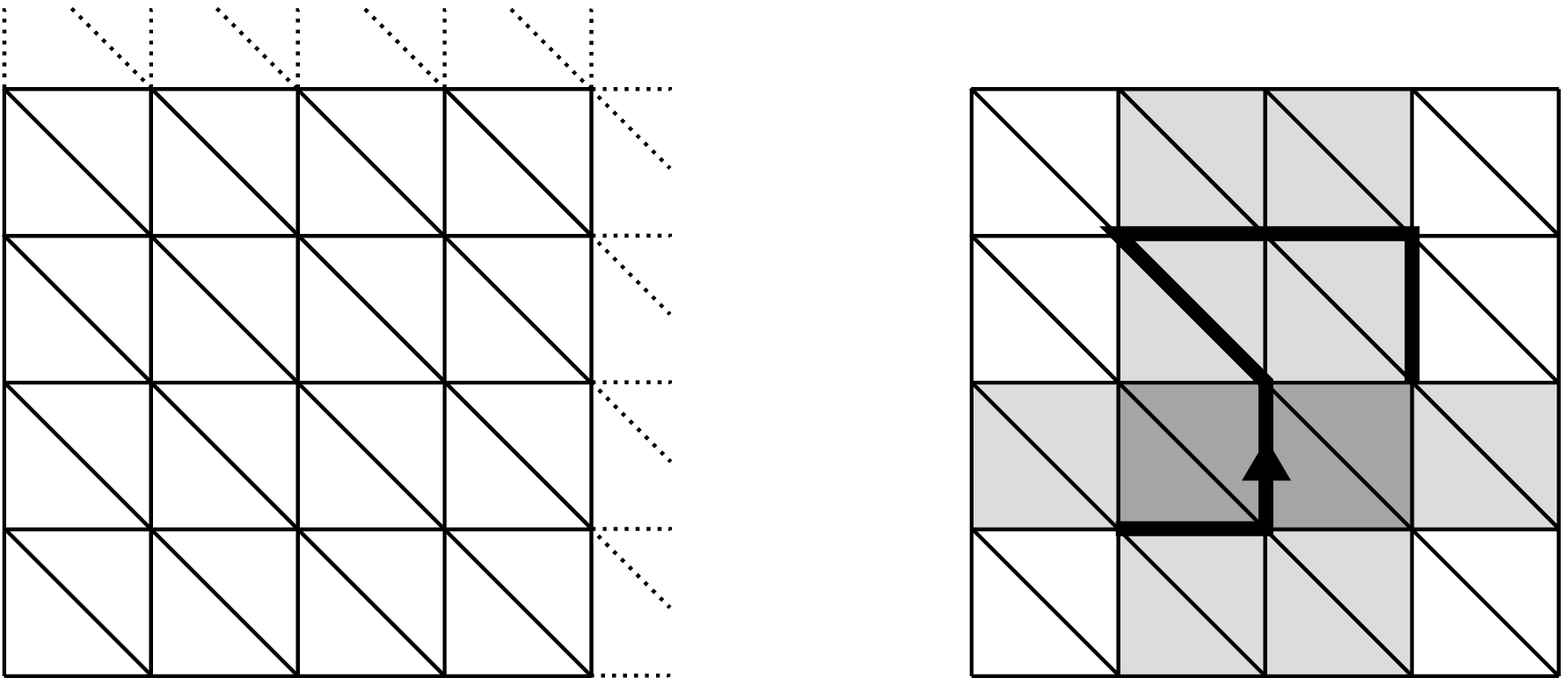
	\caption{(a) The chemical lattice for the chemical reaction network Eq.\,(\ref{eq:CN}). Horizontal edges correspond to reaction 1, vertical edges correspond to reaction 2, and diagonal edges correspond to reaction 3. (b) A path. The shaded horizontal region corresponds to $J_{\x_2} = +1$, the shaded vertical region corresponds to $J_{\x_1} = +1$, white regions correspond to vanishing currents.}
	\label{fig:CN1}
	\end{figure}

In the following we will drop all explicit time dependencies. 
We define the currents as the stochastic variables that count the net number of transitions between site $\X$ and a neighboring site, 
\bea
j_{\X+\bo{\nu}_r,\X} & := &  \# \left( \textrm{transitions from } \X \textrm{ to } \X + \bo{\nu}_r \right) \nonumber \\
& & \qquad - \# \left( \textrm{transitions from } \X + \bo{\nu}_r \textrm{ to } \X \right).
\eea
Notice that $\sum_{\X} j_{\X+\bo{\nu}_r,\X} = J_r$. Each transition decreases the Gibbs free energy of the system by an amount
\bea
f_{\X+\bo{\nu}_r,\X} & := & \ln \frac{w_{\X+\bo{\nu}_r,\X}}{w_{\X,\X+\bo{\nu}_r}},
\eea
i.e. $f_{\X+\bo{\nu}_1,\X} = \ln \y_1/(\x_1 +1)$, $f_{\X+\bo{\nu}_2,\X} =  \ln \x_2/\y_2$, $f_{\X+\bo{\nu}_3,\X} = \ln (\y_3 X_1)/(\y_4 (\x_2 +1))$.
% \end{align} 
%\begin{align}
%F_{\X+\bo{\nu}_1,\X} & = \ln \frac{\y_1 }{\x_1 +1}, &
%F_{\X+\bo{\nu}_2,\X}  & =  \ln \frac{\y_2 }{\x_2 +1}, &
%F_{\X+\bo{\nu}_3,\X}  & = \ln \frac{\y_3 X_1}{\y_2 (\x_2 +1)}. \label{eq:fex}
%\end{align} 
Notice that both the currents and the Gibbs free energy differences are antisymmetric by inversion of the orientation of the transition,
\bes
j_{\X,\X+\bo{\nu}_r} & = & - j_{\X+\bo{\nu}_r,\X} \\
f_{\X,\X+\bo{\nu}_r} & = & - f_{\X+\bo{\nu}_r,\X}.
\ees
A crucial observation is that the Gibbs free energy differences satisfy Kirchoff's Loop Law (KLL)
\bea
f_{\X,\X-\bo{\nu}_2} + f_{\X-\bo{\nu}_2,\X+\bo{\nu}_3} + f_{\X+\bo{\nu}_3,\X} = \ln \frac{\y_1 \y_3} {\y_2 \y_4}=: F \label{eq:kll}
\eea
where the {\it affinity} $F$ is the total Gibbs free energy decrease around a cyclic process that starts at $\X$ and moves by amount $\bo{\nu}_3$, then $\bo{\nu}_1$, then $\bo{\nu}_2$ to return to $\X$ by virtue of Eq.\,(\ref{eq:notind}). Quite importantly, it is peculiar to chemical networks with mass-action law that the affinity does not depend on the state $\X$ where the cycle is based, which will allow a significant simplification.

Finally we introduce the total entropy production
\bea
\Sigma : =  \sum_{\X} \sum_{r > 0} j_{\X+\bo{\nu}_r,\X} \, f_{\X+\bo{\nu}_r,\X}.
\eea
Notice that we restricted the sum to the positive verse of the reactions to avoid double-counting. This expression simplifies in view of KLL,
\bea
\Sigma
& =  &  \sum_{\X} \left[j_{\X+\bo{\nu}_3,\X} \,\left(F -  f_{\X,\X-\bo{\nu}_2} - f_{\X-\bo{\nu}_2,\X+\bo{\nu}_3}   \right) + \sum_{r =1,2} j_{\X+\bo{\nu}_r,\X} \, f_{\X+\bo{\nu}_r,\X} \right] \nonumber \\
& =  &   F J_3
+ \sum_{\X} \left[\left(j_{\X+\bo{\nu}_1,\X} + j_{\X-\bo{\nu}_3,\X}\right) f_{\X+\bo{\nu}_1,\X}   + \left(j_{\X-\bo{\nu}_2,\X} + j_{\X+\bo{\nu}_3,\X}\right) f_{\X-\bo{\nu}_2,\X}   \right] \nonumber \\
& =  &   F J_3
+ \sum_{\x_1} J_{\x_1} \ln \frac{\y_1}{\x_1 +1}
+ \sum_{\x_2} J_{\x_2} \ln \frac{\y_2}{\x_2 +1} \label{eq:SSigma}
\eea
where we introduced
\bes
J_{\x_1} := \sum_{\x_2} \left(j_{\X+\bo{\nu}_1,\X} + j_{\X-\bo{\nu}_3,\X}\right) \\
J_{\x_2} := \sum_{\x_1} \left(j_{\X-\bo{\nu}_2,\X} + j_{\X+\bo{\nu}_3,\X}\right)
\ees
respectively with the meaning of total increase of species $1$ at fixed $\x_2$ and total increase of species $2$ at fixed $\x_1$. We now introduce the second crucial ingredient, namely Kirchhoff's Current Law (KCL). Since the trajectory is continuous, the total current out of a given state visited by the trajectory must be zero, but for states $\X_0 = \X(0)$ and $\X_t = \X(t)$ that are respectively a source and a sink of a unit current. KCL can then be integrated to give
\bes
J_{\x_1} & = & \theta_{[\x_1(0),+\infty)}(\x_1) -  \theta_{[\x_1(t),+\infty)}(\x_1) \\
J_{\x_2} & = &   \theta_{[\x_2(0),+\infty)}(\x_2) - \theta_{[\x_2(t),+\infty)}(\x_2)
\ees 
where $\theta$ is the Heaviside step function on a discrete set \footnote{Defined as $\theta_{A}(b) = \sum_{a \in A} \delta_{a,b}$, for $A \subseteq \mathbb{Z}$, $\delta$ the Kroenecker delta.} (see Fig.\,\ref{fig:CN1}(b) for clarification). Employing KCL we obtain for the entropy production
\bea
\Sigma = F J_3  + \ln  \prod_{\X=\X_0+\bo{1}}^{\X_t}   \frac{\y_1} {\x_1} \frac{\y_2}{\x_2}. \label{eq:sigmaCN}
\eea

To give an interpretation of the latter term, we resort to the chemical master equation that rules the evolution of the probability of being at $\X$ at time $t$
\bea
\frac{d}{dt} P_t({\X}) = \sum_{\X,r} \Big[ w_{\X,\X+\bo{\nu}_r} P_t({\X+\bo{\nu}_r}) -  w_{\X+\bo{\nu}_r,\X} P_t({\X}) \Big] =: \mathcal{L}(\bo{\y})P_t (\X)
\eea
where on the right-hand side we introduced the generator $ \mathcal{L}({\bo{\y}})$, which of course depends on the chemostats' concentrations. The claim in Sec.\,\ref{note} is that the second term in Eq.\,(\ref{eq:sigmaCN}) should be obtained as the equilibrium distribution of the system where the third reaction is inhibited, which is achieved by setting $\y_3 = \y_4 = 0$. We then look for the solution of
\bea
\mathcal{L}(\y_1,\y_2,0,0) P^{\mathrm{eq}} = 0,
\eea
which is easily seen to be a Poissonian
\bea
P^{\mathrm{eq}}(\X) = e^{-\y_1 - \y_2} \frac{\y_1^{\x_1}}{\x_1!} \frac{\y_2^{\x_2}}{\x_2!}
\eea
satisfying detailed balance
\bea
w_{\X + \bo{\nu}_r,\X} P^{\mathrm{eq}}(\X) = w_{\X,\X + \bo{\nu}_r} P^{\mathrm{eq}}(\X + \bo{\nu}_r). 
\eea
Finally, we obtain  the desired result
\bea
\Sigma = J_3 F + \ln \frac{P^{\mathrm{eq}}(\X_t)}{P^{\mathrm{eq}}(\X_0)}.
\eea
FT's for the chemical master equation in the form of Eqs.\,(\ref{eq:psigma})-(\ref{eq:Pbo}) can be derived by standard techniques, hence supporting the result that exact FT's for the currents hold when the initial state is sampled from the equilibrium ensemble obtained by disconnecting the mechanisms that drive the system to nonequilibrium. In the next sections we will provide the full theory and in Sec. \ref{CN1} we come back to this example to discuss how the general theory allows is to generalize these observations to arbitrary chemical networks.

\section{\label{tools} Tools: Cycle and cocycles in network thermodynamics}

We will be involved with continuous-time Markov jump process on a finite state space. The state space of the system can be viewed as an oriented graph, the trajectory followed by the jump process as a sequence of oriented edges connecting vertices. All thermodynamic observables associated to the trajectory (current, free energy increase etc.) are weights assigned to every edge of the graph, antisymmetric by inversion of the orientation of the edge. In this section we briefly refresh the ensuing algebraic graph-theoretical picture; a broader treatment can be found in Ref. \cite{polettini2}.

\subsection{Cycle/cocycle decomposition of a graph}

The state space of the system is a connected oriented graph $G = (X,E)$ (without loops, allowing multiple edges) with oriented edges $e \in E$ connecting distinct vertices $x \in X$. Let $|E|$ be the number of edges and $|X|$ that of vertices. The orientation is arbitrary, by $-e$ we represent the inverse orientation of an edge. The graph is completely characterized by the $|E| \times |X|$ matrix $\partial$ prescribing the incidence relations between edges and vertices:
\bea
\partial_{x,e} = \left\{
\ba{ll} +1 , & \mathrm{if}\; \stackrel{e}{\longrightarrow} x  \\
-1 , & \mathrm{if}\; \stackrel{e}{\longleftarrow} x \\
0,  & \mathrm{otherwise}
\ea \right. .
\eea
We will make constant reference to the following example:
\bea
\ba{c}\xymatrix{ x_1  \ar[r]^{e_1} & x_2 \ar[d]^{e_2} \\ 
x_4 \ar[ur]_{e_5} \ar[u]^{e_4} & x_3 \ar[l]^{e_3} } \ea, \qquad \partial = \left(\ba{ccccc}
-1 & 0 & 0 & +1 & 0 \\
+1& -1& 0 & 0 & +1 \\
0 & +1 & -1 & 0 & 0 \\
0 & 0 & +1 & -1 & -1
\ea\right). \eea
Real combinations of edges are denoted by a vector in Dirac notation $\ket{\cdot} \in \mathbb{R}^E$. We also introduce the transpose vector $\bra{\cdot}$ and the scalar product $\braket{\cdot}{\cdot}$.

Cycles are, as intuitive, successions of oriented edges (a tail for each tip, at every vertex). Cycles are algebraically characterized as integer right null vectors of the incidence matrix,
\bea
\partial \, \ket{c} = 0.
\eea 
Therefore they form a vector space. A preferred basis of cycles can be constructed by a standard procedure that was employed by Schnakenberg for the analysis of network thermodynamics \cite{schnak}. We briefly refresh it. A {\it spanning tree} $T$ is a maximal set of (unoriented) edges that contains no cycles. We choose one such arbitrary spanning tree,
\bea
\ba{c}\xymatrix{ \ar@{-}[r]^{e_1} &  \ar@{-}[d]^{e_2} 
\\    \ar@{.}[ur]_{e_5}  \ar@{.}[u]^{e_4}  & \ar@{-}[l]^{e_3} }\ea. \eea
The choice of a spanning tree is arbitrary from a mathematical point, while physically it corresponds to the choice of a different set of relevant observables. An important property of spanning trees is that there exists a unique oriented path $\gamma_{xx'}$ connecting any vertex $x'$ to any other vertex $x$ of the graph belonging to the spanning tree. 

Edges not belonging to the spanning (dotted, above) are called the {\it chords} $e_\alpha$. Their number is given by Euler's formula $|C| = |E| - |X| +1$. Adding chord $e_\alpha$ to the spanning tree identifies a unique cycle $c_\alpha$ with orientation along the verse of the chord:
		\bea
		c_4 = \!\!
		\ba{c}\xymatrix{ \ar@{->}[r]  \ar@{<-}[d] &  \ar@{->}[d] 
		\\   \ar@{->}[u] &  \ar@{->}[l]  }\ea,
		\qquad
		c_5 =  \!\!
		\ba{c}\xymatrix{ & \ar@{->}[d] 
		\\   \ar@{->}[ur]  & \ar@{->}[l] }\ea.
		\eea
It can be proven that the set of cycles so generated is a basis for the null space of the incidence matrix \cite{nakanishi}.

Orthogonal to the set of  cycles is the set of {\it cocycles} (or cuts), generated by the corresponding {\it cochords}. A cochord is an edge $e^\ast_\mu$ belonging to the spanning tree. Their number is $|E|-|C| = |V|-1$. Removing cochord $e^\ast_\mu$ from a spanning tree disconnects the graph into two basins. The set of edges that connect one basin to the other, oriented in the verse of the generating cochord, is a cocycle:
\bea
		c^\ast_1 = \!\! \ba{c}\xymatrix{ \bullet \ar[r]  \ar[d] & \circ
		\\ \circ  & \circ \ar@{.}[u] \ar@{.}[l]  }\ea, \quad
		c^\ast_2 = \!\!  \ba{c}\xymatrix{\bullet \ar@{->}[d]  \ar@{.}[r] & \bullet
		\\  \circ \ar@{<-}[ur] \ar@{.}[r] & \circ \ar@{<-}[u] }\ea, \quad
		c^\ast_3 = \!\!  \ba{c}\xymatrix{\bullet \ar@{.}[r] & \bullet
		\\ \ar@{<-}[ur]  \ar@{<-}[u]  \circ & \ar@{->}[l] \ar@{.}[u]  \bullet  } \ea. \label{eq:cocyclefig}
\eea
In this example, vertices in the source basins are disks, in the target basins are circles, edges of the spanning tree that connect them are dotted.

We can now give a vector representation of chords, cycles, corchords, and cocycles as linear combinations of edges of the graph. We denote them respectively $\ket{e_\alpha}, \ket{c_\alpha}$, $\ket{e_\mu^\ast}$, $\ket{c_\mu^\ast}$. In our example, we have
\bea
\ket{c^\ast_1} = \left(\ba{c} +1 \\ 0 \\ 0 \\ -1 \\ 0\ea\right), \quad
\ket{c^\ast_2} = \left(\ba{c} 0 \\ +1 \\ 0 \\ -1 \\ -1 \ea\right), \quad
\ket{c^\ast_3} = \left(\ba{c} 0 \\ 0 \\ +1 \\ -1 \\ -1 \ea\right), \nonumber  \\
\ket{c_4} = \left(\ba{c} 1 \\ 1 \\ 1 \\ 1 \\ 0 \ea\right), \quad
\ket{c_5} = \left(\ba{c} 0 \\ 1 \\ 1 \\ 0 \\ 1\ea\right).
\eea

A crucial result proven in Ref.\,\cite{polettini2} is that the identity over the edge space can be decomposed as
\bea
I = \sum_\alpha \ket{c_\alpha} \bra{e_\alpha} +  \sum_\mu \ket{e^\ast_\mu} \bra{c^\ast_\mu}. \label{eq:I}
\eea
where $\ket{\,\cdot\,} \bra{\,\cdot\,}$ denotes the outer product of two vectors, yielding an $|E| \times |E|$ matrix.  We will repeatedly employ this identity in the following sections. As a side comment, $P = \sum_\alpha \ket{c_\alpha} \bra{e_\alpha}$ and $P^\ast = \sum_\mu \ket{e^\ast_\mu} \bra{c^\ast_\mu}$ are oblique complementary projectors, $P^2 = P$, ${P^\ast}^2 = P^\ast$, $PP^\ast = P^\ast P = 0$, which gives rise to an elegant formulation of network thermodynamics based on projectors.

\subsection{Tidal and cycle currents and their dual variables}

In network thermodynamics one assigns two observables to each oriented edge, the current $j_e$ and its conjugate force $f_e$. They are required to be antisymmetric by edge inversion, $j_{-e} = - j_{e}$, $f_{-e} = - f_{e}$. We collect their values in two vectors $\ket{j}$ and $\ket{f}$. The  {\it entropy production} is the bilinear form
\bea
\Sigma = \braket{j}{f} = \sum_e f_e j_e.
\eea
We immediately apply the identity decomposition Eq.\,(\ref{eq:I}) to the currents to obtain 
\bea
\ket{j} & = & \sum_\alpha \ket{c_\alpha} \braket{e_\alpha}{j} +  \sum_\mu \ket{e^\ast_\mu} \braket{c^\ast_\mu}{j}. \label{eq:currproj} \\
& =: & \sum_\alpha  J_\alpha \ket{c_\alpha} +  \sum_\mu  J_\mu \ket{e^\ast_\mu}. \label{eq:currproj}
\eea
The second line defines the {\it cycle currents} $J_\alpha$ and the {\it tidal currents} $J_\mu$. The former are well-known from Schnakenberg's analysis. They quantify the cycling of a process. The latter give the total flux from a set of source vertices to a set of target vertices, as visualized in Eq.\,(\ref{eq:cocyclefig}). This decomposition is somewhat analogous to the Helmholtz decomposition of a vector field into a curl and a gradient (modulo a harmonic term).

Now, plugging Eq.\,(\ref{eq:currproj}) into the entropy production we obtain
\bea
\Sigma & = & \sum_\alpha  J_\alpha \braket{f}{c_\alpha} +  \sum_\mu  J_\mu \braket{f}{e^\ast_\mu} \\
& = & \sum_\alpha  J_\alpha F_\alpha +  \sum_\mu  J_\mu F_\mu \label{eq:epcc}
\eea
where we introduced the {\it affinities} $F_\alpha$ as observables conjugate to the cycle currents, and the {\it potential drops} $F_\mu$ as observables conjugate to the tidal currents.

\section{\label{results} Results: Fluctuation theorems for the currents}

\subsection{Transient FT for joint tidal and cycle currents}

We consider a continuous-time Markov jump process $(\bo{x},\bo{\tau})$, starting at state $x_0$ and performing $n$ transitions in time $t$ to state $x_t$. The rate of a jump from state $x$ to $x'$ is $w_{x'x}$. The process visits state $x_i$ for an interval $\tau_i$ before jumping to state $x_{i+1}$, up to time $t = \tau_0 + \ldots + \tau_n$. The joint probability density of the $n$ states visited by the trajectory is given by
\bea
P_{n,t}(\bo{x},\bo{\tau}) =  \delta\Big( t - \sum_{i = 0}^n \tau_i \Big)e^{- w_{x_n} \tau_n}  \prod_{i = 0}^{n-1} w_{x_{i+1},x_i} e^{- w_{x_i} \tau_i}  ,
\eea
where $\delta(\cdot)$ is the Dirac delta and $w_x = \sum_{x'} w_{x'x}$. The marginal probability density for the states is given by
\bea
P_{n,t}(\bo{x}) = \int\!\ldots \!\int_0^\infty   d\bo{\tau} \,P_{n,t}(\bo{x},\bo{\tau}) = Q_{n,t}(\bo{x}) \prod_{i = 0}^{n-1} w_{x_{i+1},x_i} 
\label{eq:marginalt}
\eea
where
\bea
Q_{n,t}(\bo{x}) :=   \int\!\ldots \!\int_0^\infty   d\bo{\tau} \,  \delta\Big( t - \sum_{i = 0}^n \tau_i \Big)  \prod_{i = 0}^{n} e^{- w_{x_n} \tau_n}.
\eea
The actual dependence on $t$ is difficult to compute and not relevant for what follows.

We define the time-reversed process as that process $(\bo{x}^\dagger,\bo{\tau}^\dagger)$ where the succession of states and time intervals are inverted, $x^\dagger_i = x_{n-i}$ and $\tau^\dagger_i = \tau_{n-i}$. Notice that for the time-reversed process we have
\bea
Q_{n,t}(\bo{x}) = Q_{n,t}(\bo{x}^\dagger).
\eea
As a consequence, the following fluctuation relation between forward and backward successions of states holds
\bea
P_{n,t}(\bo{x})  = P_{n,t}(\bo{x}^\dagger) \prod_i \frac{w_{x_{i+1},x_i}}{w_{x_i,x_{i+1}}} . \label{eq:ftx}
\eea

The above expression can be further marginalized. We define the (time-integrated) edge current along $x\gets x'$ as a stochastic variable counting the net number of transitions from $x'$ to $x$, 
\bea
j_{xx'}(\bo{x}) := \sum_{i} \left(\delta_{x_{i+1},x} \, \delta_{x_i,x'} - \delta_{x_{i+1},x'} \, \delta_{x_i,x} \right).
\eea
It satisfies the antisymmetry relations $j_{x'x}(\bo{x})  = - j_{xx'}(\bo{x}) $ and $j_{xx'}(\bo{x}) = - j_{xx'}(\bo{x}^\dagger)$. Eq.\,(\ref{eq:ftx}) can then be written in terms of the currents as follows
\bea
P_{n,t}(\bo{x})  % & = &  \prod_{x < x'} \left( \frac{w_{xx'}}{w_{x'x}} \right)^{j_{xx'}} \, P_{t}(\bo{x}^\dagger|n) \\
= e^{\braket{f}{j(\bo{x})}} \, P_{n,t}(\bo{x}^\dagger)  
\eea
where the entries of $\ket{f}$ are the thermodynamic forces $f_{xx'} := \ln (w_{xx'}/w_{x'x})$.
We can then finally marginalize for the currents taking values $\ket{j}$. It must be here noted that the expressions of the current and of the probability measure are conditioned to a fixed total number of transitions $n$. Although, experimentally one usually has access to the total number of transitions between two states irrespective of the total number of transitions that the trajectory performs. The probability of observational values of the currents $\ket{j}$ up to time $t$ is given by
\bea
P_{t}(\ket{j}) = \sum_{n=1}^\infty P_{n,t}(\ket{j})P_{t}(n) 
\eea 
where $P_{t}(n)$ is the probability that a total number of transitions $n$ occurs in time $t$. Since the time-reversed process performs the same number of jumps, we do not need to compute it, and we obtain
\bea
P_{t}(\ket{j}) = e^{\braket{f}{j}} P_{t}(-\ket{j}).
\eea

We can now apply the cycle/cocycle decomposition exposed in Sec.\,\ref{tools}. To do this, we should first identify a spanning tree of the graph such that the chord currents are currents of physical relevance to the specific model at hand. We can then define stochastic cycle and tidal currents 
\bes
J_\alpha(\bo{x}) & := & \braket{e_\alpha}{j(\bo{x})} \\
J_\mu(\bo{x}) & := &  \braket{c_\mu}{j(\bo{x})}.
\ees
The first counts the number of times the $\alpha$-th cycle is enclosed, the second counts the number of times the process jumps from the source to the target basin of the $\mu$-th cut. Since cycle and tidal currents are one-to-one to the edge currents, by a simple coordinate transformation (which can be proven, but here is irrelevant, to have unit Jacobian) we can move to the probability of the former, which by Eq.(\ref{eq:epcc}) obeys the joint FT
\bea
\frac{P_{t}(J_\alpha,J_\mu)}{P_{t}(-J_\alpha,-J_\mu)} = \exp\left(\sum_\alpha F_\alpha J_\alpha + \sum_\mu F_\mu J_\mu \right). \label{eq:jft}
\eea
This equation generalizes the result of Andrieux and Gaspard to finite-times. An important observation is that we do not need to condition this FT to an initial and a final state, since conditioning is implicit. In fact, knowledge of the tidal currents implies knowledge of the initial and final states.

\subsection{Asymptotic FT for the cycle currents}

We now focus on the tidal term. An important fact is that a Markov jump process on a graph is continuous, i.e. it can be drawn without lifting the pencil. As an important implication, tidal currents can only take values in $\{-1,0,1\}$, while cycle currents take values in $\mathbb{Z}$. Intuitively, while a process can wind arbitrarily many times around a cycle in a preferential direction, the only way to increase a tidal current is to move from the source to the target basin of the cocycle, after which by continuity only the inverse can occur, restoring the tidal current to its initial value. In fact, orienting all edges of the graph in such a way that the initial state $x_0$ is in the source basin of all cocycles, then tidal currents can only take values in $\{0,1\}$.  

Let $\bra{\partial_x}$ be the rows of the incidence matrix. Then by continuity of the trajectory
\bea
\braket{\partial_x}{j(\bo{x})} = \delta_{x,x_t} -  \delta_{x,x_0}.
\eea
This also shows that knowledge of the complete set of currents retains the infomation about the initial and final states, that is, {\it de facto} the FT Eq.\,(\ref{eq:jft}) is conditioned to its boundary states. Since by definition the kernel of $\partial$ is the cycle space, then the row space of the incidence matrix spans the cocycle space. Then there exists a linear transformation $M$ such that $\bra{c_\mu} = \sum_x M_{\mu,x} \bra{\partial_x}$, given by
\bea
M_{\mu,x} = \left\{\ba{ll} - 1/2 , & x \in \mathrm{source}~ c_\mu \\
+ 1/2 , & x \in \mathrm{target}~ c_\mu \ea 
\right. . \label{eq:top}
\eea
The $\pm 1/2$ terms are given by the fact that the rows of the incidence matrix are not linearly independent, and therefore one has to adjust a double counting. Then
\bea
J_\mu(\bo{x}) = \sum_x M_{\mu,x} \braket{\partial_x}{j(\bo{x})} = M_{\mu,x_t} - M_{\mu,x_0}, \label{eq:01}
\eea
which is $0$ if both $x_0$ and $x_t$ are in the target or in the source, $+1$ if $x_0$ is in the source and $x_t$ in the target, and $-1$ vice versa.

It follows from this discussion that tidal currents are bounded, while cycle currents typically increase with time according to
\bea
J_\alpha(\bo{x}) \asymp t \, \j_\alpha(\bo{x}),
\eea
where $\j_\alpha$ is the current per time, and when referred to a stochastic variable $\asymp$ means asymptotically, almost surely. Then in the long time limit the asymptotic FT of Andrieux and Gaspard is obtained
\bea
\lim_{t\to+\infty}\frac{1}{t} \ln \frac{P_{t}(t\,\j_\alpha)}{P_{t}(-t\,\j_\alpha)} \asymp \sum_\alpha F_\alpha \, \j_\alpha . \label{eq:jfta}
\eea 

Unless a restoring force intervenes (e.g. periodic driving, time-dependent protocols etc.), tidal forces are doomed to disappear. Indeed, their effect is so week that they do not affect any statistical property of the currents \cite{artur}.

\subsection{\label{uncon}Unconditional transient FT for the cycle currents (without ensembles)}

Let us define a function over the vertices
\bea
\Phi_x := - \sum_\mu F_\mu M_{\mu,x}. \label{eq:pot}
\eea
After Eq.\,(\ref{eq:01}) we obtain
\bea
\sum_\mu F_\mu J_\mu(\bo{x}) = \Phi_{x_0} - \Phi_{x_t},
\eea
which means that the tidal contribution is a state function. An intuitive way to picture this is the following. Suppose the trajectory moves from state $x_0$ to $x_t$ along the spanning tree. Then the tidal term is increased by the potential drops within the tree and the cycle term is untouched. Now, instead, suppose that $x_0$ and $x_t$ are connected by a chord, and that the trajectory travels along that chord. Then, one will account one full cycle and consequently will have to subtract terms from the tidal accounting. As far as the cocycle term is concerned, the result of these two operations is the same, that is, the tidal term only cares about where the trajectory is and not how it got there, because every time a cycle is enclosed that contribution is thrown in the cycle term.

Then, we can express the joint FT in terms of the cycle currents, conditioned to the boundary states:
\bea
\frac{P_t(J_\alpha|x_0)}{P_t(- J_\alpha|x_t)} = \exp \left( \sum_\alpha F_\alpha  J_\alpha + \Phi_{x_0} - \Phi_{x_t}\right). \label{eq:bb} 
\eea

Now suppose the initial state $x_0$ is sampled with probability $P_0(x_0)$, and that the initial state of the time-reversed processes is sampled with probability $P_t(x_t)$. The choice
\bea
P_0(x) = P_t(x) = Z^{-1} e^{-\Phi_x}
\eea
clearly de-conditions the above expression with respect to the boundary states, which can then be marginalized yielding the finite time FT for the cycle currents, with given initial state
\bea
\frac{P_t(J_\alpha)}{P_t(- J_\alpha)} = \exp \sum_\alpha F_\alpha J_\alpha .
\eea

Finally, let us give a clear interpretation of the special distribution from which boundary states must be sampled to attain an exact finite-time FT. By definition $F_\mu$ is the potential drop across the generating cochord, which belongs to the spanning tree. Fixing a reference state $\bar{x}$, let $\ket{\gamma_{x\bar{x}}}$ be the vector representative of the unique oriented path in the spanning tree that connects $\bar{x}$ to $x$. Then one has
\bea
\Phi_x = \braket{\gamma_{x\bar{x}}}{f}.
\eea
It is then well known that the state $Z^{-1} e^{-\Phi_x}$ is the equilibrium steady state of the network where the chords are completely removed. Changing reference state $\bar{x}$ amounts to shifting the potential $\Phi$ by a constant ground value.

Let us recapitulate this important message. Consider a continuous-time Markov jump process on a graph. Choose a spanning tree of the graph. The criterium is that the currents flowing across the chords  (i.e. edges not belonging to the spanning tree) should be of particular physical relevance.  Then, such cycle currents satisfy an exact transient fluctuation relation if the processes are sampled from the equilibrium ensemble reached by the Markov process with all rates along chords set to zero.

Finally, as is well-known equilibrium ensembles can be obtained by a maximum entropy procedure \cite{jaynes} with suitable constraints that incorporate the information available about the system before the experiment is conducted, which is used to build up a prior probability (on the role of priors in nonequilibrium statistical mechanics at a foundational level, see Refs.\,\cite{prior,gauge} by one of the authors). Then, it is interesting to note that the ensemble that needs to be prepared for an observation of the FT at all times is precisely the maximum entropy ensemble (the state of lowest information) with respect to the experimental apparatus that is going to measure the currents. The initial ensemble is dictated uniquely by the topology of the graph, expressed by Eq.\,(\ref{eq:top}) and by the potential Eq.\,(\ref{eq:pot}) whose average plays the role of the maximum entropy constraint according to the theory pioneered by Jaynes \cite{jaynes}. 

\section{\label{CN2}Example: network of chemical reactions revisited}

\begin{figure}[t]
  \centering
 \def\svgwidth{400pt}
 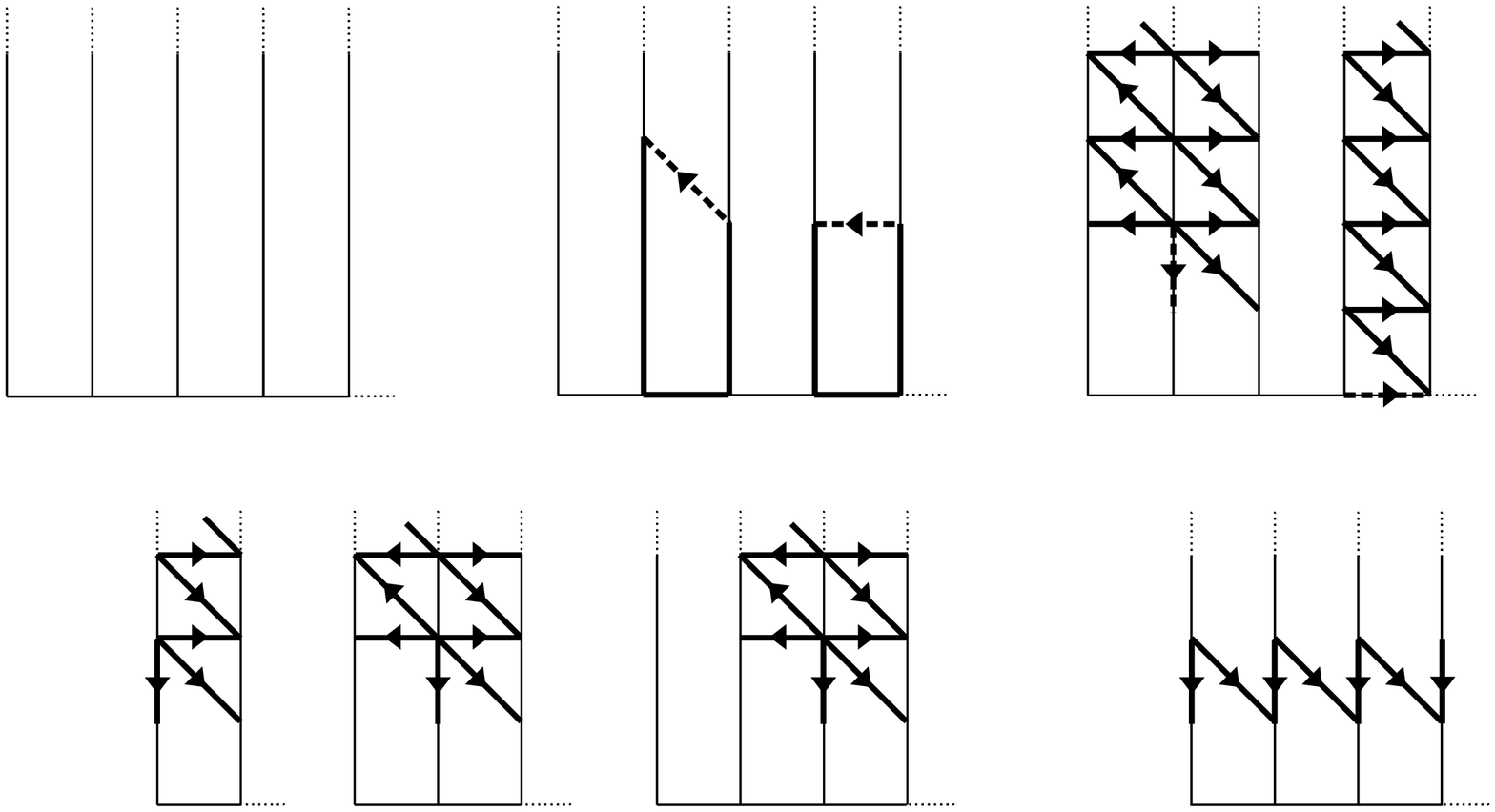
  \caption{(a) An infinite spanning tree (b) Two cycles; the generating chords are dashed. (c) Two cocycles; the generating cochords are dashed. Cocycle $c^\ast_2$ is generated by a vertical cochord, cocycle $c^\ast_1$ is generated by a horizontal cochord. (d) Linear combination of cocycles of type $c^\ast_2$ at given $\x_2$.}
  \label{fig:CN}
\end{figure}

The graph-theoretical method exposed in Secs.\,\ref{tools} and \ref{results} can be fruitfully applied to the chemical network analyzed in Sec.\,\ref{CN1}, conjecturing that all results can be extended to the infinite case in some mathematically rigorous way. The chemical lattice admits an infinite number of spanning trees, most of which have no regularity. We choose the {\it comb} depicted in Fig.\,\ref{fig:CN}(a), consisting of the edges along the $\x_1$ axis and of all the vertical edges. With reference to Fig.\,\ref{fig:CN}(b), there are two kinds of cycles: Cycles of kind $c_0$ generated by chords $\X \to \X+\bo{\nu}_1$ have null affinity; Cycles of kind $c_3$ generated by chords  $\X \to \X+\bo{\nu}_3$ have affinity $F =  \ln \frac{\y_1 \y_3} {\y_2 \y_4}$. Hence the cycle term reads
\bea
\sum_\alpha F_\alpha J_\alpha = F \sum_{\X} j_{\X+\bo{\nu}_3,\X}  =  F J_3
\eea
yielding the first term in Eq.\,(\ref{eq:sigmaCN}).

There are two types of cocycles. Horizontal cochords $(\x_1,0)\to (\x_1+1,0)$  carrying potential drop $\ln \y_1/(\x_1+1)$ generate cocycles of type $c^\ast_1$ in Fig.\,\ref{fig:CN}(c) with current $J_{\x_1}$. As regards the vertical set of cochords of type $\X\to \X+\bo{\nu}_2$, notice that all those that are based at the same $\x_2$ carry the same potential drop $\ln \x_2/\y_2$. Then a resummation occurs, as depicted in Fig.\,\ref{fig:CN}(d), and one obtains an effective cocyle carrying current $J_{\x_2}$. We then obtain
\bea
\sum_\mu F_\mu^\ast J^\ast_\mu = \sum_{\x_1}  J_{\x_1} \ln \frac{\y_1}{\x_1+1} +  \sum_{\x_2}  J_{\x_2} \ln \frac{\y_2}{\x_2+1} 
\eea
which is the second term in Eq.\,(\ref{eq:SSigma}).

Finally, the initial equilibrium ensemble that makes the finite-time FT for the currents hold is the steady state of a Markov process occurring on the comb, which is obtained by eliminating reaction 3 from the reaction scheme. It is interesting to note that, due to the fact that the affinities of type 0 in Fig.\,(\ref{fig:CN}) all vanish, any spanning tree that only consists of reaction steps of kind 1 and 2 will give rise to the same initial ensemble. Hence, while in principle the choice of the spanning tree affects the FT, chemical networks enjoy certain regularity properties that boil down the great generality of Schnakenberg's analysis to actual physical currents. In the specific case of chemical networks, this possibility is granted by the mass-action law and by the fact that the topology of the chemical lattice in Fig.\,\ref{fig:CN}(a) is simply obtained by shifting and reproducing the chemical reaction network in Eq.\,(\ref{eq:cn}). While we postpone a full discussion of chemical networks to a future publication, it is interesting to note that not all chemical reaction schemes allow for such great simplification depending on certain topological properties related to the concept of {\it deficiency} of the network \cite{deficiency}.

\section{\label{conclusions}Conclusions}

In this paper we collected several results about finite-time FT's for the currents for stationary Markov jump processes on a finite state space, giving a unified framework based on certain algebraic graph-theoretical techniques that allow to decompose any thermodynamic observable in terms of cycle and cocycle observables, by virtue of the fundamental identity Eq.\,(\ref{eq:I}).  In particular, we generalized the result of Andrieux and Gaspard \cite{andrieux} for the so-called Schnakeberg currents to finite times, both by direct manipulation of the probability density function of Markov jump trajectories, and by the generating function approach exposed in \ref{appendix}.

One major limitation of our results that calls for further generalization is the requirement that transition rates are time-independent. Indeed, the FT without ensemble quantities discussed by Bulnes-Cuetara et al. \cite{cuetara} was formulated for time-dependent protocols. We mention, without further discussion, that a full generalization of their result to arbitrary Markov processes on finite state spaces in terms of cycles and cocycles is significantly more complicated. The resulting expressions defy a clear physical interpretation. Partial results can be obtained under more restrictive assumptions, e.g. that the affinities are constant in time. Furthermore, as regards linear chemical networks we point out that there exists a finite-time FT with the initial state sampled from the steady nonequilibrium ensemble, with a time-dependent effective affinity \cite{andrieux3}. We leave these issues and the treatment of general chemical reaction networks to future inquiry.

\appendix

\section*{Acknowledgements}

We thank G. Bulnes-Cuetara for discussion and A. Wachtel for comments on the manuscript. The research was supported by the National Research Fund Luxembourg in the frame of project FNR/A11/02 and of Postdoc Grant 5856127.

\appendix

\section{\label{appendix}Generating function approach}

As an {\it addendum}, we will prove the ``initial ensemble'' FT exposed in Sec.\,\ref{uncon} by the commonly employed method of the generating function, generalizing the treatment explored by Bulnes-Cuetara in Ref.\,\cite{gregthesis} and by Andrieux et al. in Ref.\,\cite{agmt}.

Let $\bo{\lambda} = (\lambda_\alpha)_\alpha$ be a set of counting fields defined on the chords of the graph and $\bo{1} = (1,\ldots,1)$ the unit vector of length $|V|$. It is well-known \cite{andrieux} that the moment generating function for the cycle currents is given by
\bea
Z(\bo{\lambda},t) = \sum_x P_x(\bo{\lambda},t) = \bo{1} \cdot P(\bo{\lambda},t), \label{eq:Z}
\eea
where $P(\bo{\lambda},t)$ evolves by the Feynman-Kac type of equation
\bea
\frac{d}{dt} P(\bo{\lambda},t) = \mathcal{L}(\bo{\lambda}) P(\bo{\lambda},t). \label{eq:fk}
\eea
Here, $\mathcal{L}(\bo{\lambda})$ is  the {\it tilted} generator with entries
\bea
\mathcal{L}_{xx'}(\bo{\lambda})  = \left\{ \ba{ll}  - w_x, &  \textrm{if }  x = x' \\ w_{xx'} e^{\lambda_{\alpha}}, & \textrm{if } x\neq x',  x \gets x' = e_\alpha \\
w_{xx'} e^{-\lambda_{\alpha}},  & \textrm{if } x\neq x',  x \to x' = e_\alpha \\
w_{xx'} &   \textrm{otherwise}  \ea \right. 
\eea
and the initial condition is given by
\bea
P(\bo{\lambda},0) = P(\bo{0},0),
\eea
where $P(\bo{0},0)$ is the initial probability density over states. It is important that $P(\bo{\lambda},0)$ does not depend on $\bo{\lambda}$. Physically, this reflects the fact that the preparation of the system cannot depend on the output of the counting experiment. 

The tilted generator obeys a crucial time-reversal symmetry relation. Let us consider the generator $\mathcal{L}(\bo{F} - \bo{\lambda})^T$ with entries
\bea
[\mathcal{L}(\bo{F} - \bo{\lambda})^T]_{xx'} = \left\{ \ba{ll}  - w_x, &  \textrm{if }  x = x' \\ w_{x'x} e^{F_\alpha-\lambda_{\alpha}}, & \textrm{if } x'\neq x,  x' \gets x = e_\alpha \\
w_{x'x} e^{\lambda_\alpha-F_\alpha},  & \textrm{if } x'\neq x,  x' \to x = e_\alpha \\
w_{x'x} &   \textrm{otherwise}  \ea \right. .
\eea
By definition, the $\alpha$-th affinity is the circulation of the force around a cycle comprising chord $e_\alpha = x \gets x'$ and the unique path $\gamma_{x'x}$ that is internal to the spanning tree $T$ and that goes from state $x$ to state $x'$. Then,
\bea
F_\alpha = \ln\frac{w_{xx'}}{w_{x'x}} + \braket{\gamma_{x'x}}{f} = \ln\frac{w_{xx'}}{w_{x'x}} + \Phi_{x'} - \Phi_{x}, \qquad  x \gets x' = e_\alpha .
\eea
Moreover, notice that for all edges internal to the spanning tree
\bea
\Phi_{x'} - \Phi_{x} = \ln \frac{w_{x'x}}{w_{xx'}}, \qquad   x \gets x'  \in T.
\eea
We then obtain
 \bea
[\mathcal{L}(\bo{F} - \bo{\lambda})^T]_{xx'} = \left\{ \ba{ll}  - w_x, &  \textrm{if }  x = x' \\ w_{x'x} e^{\Phi_{x'} - \Phi_{x}-\lambda_{\alpha}}, & \textrm{if } x'\neq x,  x' \gets x = e_\alpha \\
w_{x'x} e^{\Phi_{x'} - \Phi_{x}+\lambda_{\alpha}}, & \textrm{if } x'\neq x,  x' \to x = e_\alpha \\
w_{xx'} e^{\Phi_{x'} - \Phi_{x}} &   \textrm{otherwise}  \ea \right.
\eea
yielding the symmetry relation
\bea
\mathcal{L}(\bo{F} - \bo{\lambda})^T = \Theta  \mathcal{L}(\bo{\lambda}) \Theta^{-1},
\eea
where $\Theta  = \mathrm{diag}  \left(  e^{-\Phi_{x}}/Z \right)_x$, $Z= \sum_x e^{-\Phi_{x}}$ being the normalization factor. 

Finally, we can go back to the moment generating function. Integrating Eq.\,(\ref{eq:fk}), in view of Eq.\,(\ref{eq:Z}), we obtain
\bea
Z(\bo{\lambda},t) = \bo{1} \cdot e^{t \mathcal{L}(\bo{\lambda})}   P(\bo{0},0) = \bo{1} \cdot  \Theta^{-1}   e^{t \mathcal{L}(\bo{F} - \bo{\lambda})^T } \Theta P(\bo{0},0).
\eea
Let us also consider
\bea
Z(\bo{F} - \bo{\lambda},t) = \bo{1} \cdot  e^{t \mathcal{L}(\bo{F} - \bo{\lambda})} P(\bo{0},0).
\eea
In general, the two are not related unless
\bea
P(\bo{0},0) = \Theta^{-1} \bo{1}, \label{eq:incon}
\eea
in which case
\bea
Z(\bo{\lambda},t) = Z(\bo{F} - \bo{\lambda},t). \label{eq:duft}
\eea
Eq.\,(\ref{eq:incon}) is nothing but the requirement that the initial state is the equilibrium state described in Sec.\,\ref{uncon}, Eq.\,(\ref{eq:duft}) is well-known to imply the fluctuation relation when moving from the generating function picture to the probability density picture.


\section*{Bibliography}

\begin{thebibliography}{20}

\bibitem{bochkov} Bochkov G. N. and Kuzovlev Y. E., {\it Nonlinear  fluctuation-dissipation relations and stochastic models in nonequilibrium thermodynamics: I. Generalized fluctuation-dissipation theorem}, 1981 Physica A {\bf 106}, 443; {\it Nonlinear fluctuation-dissipation relations and stochastic models in nonequilibrium thermodynamics: II. Kinetic potential and variational principles for nonlinear irreversible processes}, 1981 Physica A {\bf 106}, 480.

\bibitem{kurchan} Kurchan J., {\it Fluctuation theorem for stochastic dynamics}, 1998 J. Phys. A.: Math. Gen. {\bf 31}, 3719.

\bibitem{maes} Maes C., {\it The fluctuation theorem as a Gibbs property}, 1999 J. Stat. Phys. {\bf 95}, 367.

\bibitem{lebowitz} Lebowitz J. L. and Spohn H., {\it A Gallavotti-Cohen-Type Symmetry in the Large Deviation Functional for Stochastic Dynamics}, 1999 J. Stat. Phys. {\bf 95}, 333.

\bibitem{kalpa} Kalpazidou S., 1995 {\it Cycle representations of Markov processes} (Berlin: Springer).

\bibitem{hill} Hill T. L., 2005 {\it Free Energy Transduction and Biochemical Cycle Kinetics} (New York: Dover).

\bibitem{schnak} Schnakenberg J., {\it Network theory of microscopic and macroscopic behavior of master equation systems}, 1976 Rev. Mod. Phys. {\bf 48}, 571.   

\bibitem{polettini1} Polettini M., {\it Macroscopic constraints for the minimum entropy production principle}, 2011 Phys. Rev. E {\bf 84}, 051117. 

\bibitem{andrieux} Andrieux D.  and  Gaspard P., {\it Fluctuation theorem for currents and Schnakenberg network theory}, 2007 J. Stat. Phys. \textbf{127}, 107.

\bibitem{andrieux2} Andrieux D. and Gaspard P., {\it Fluctuation theorem and Onsager reciprocity relations}, 2004 J. Chem. Phys. {\bf 121}, 6167.

\bibitem{artur} Wachtel A., Vollmer J. and Altaner B., {\it Determining the Statistics of Fluctuating Currents: General Markovian Dynamics and its Application to Motor Proteins}, 2014 arXiv:1407.2065.

\bibitem{faggionato} Faggionato A. and Di Pietro D., {\it Gallavotti-Cohen-Type Symmetry Related to Cycle Decompositions for Markov Chains and Biochemical Applications}, 2011 J. Stat. Phys \textbf{143}, 11.

\bibitem{bertini} Bertini L., Faggionato A., D. Gabrielli D., {\it Flows, currents, and cycles for Markov Chains: large deviation asymptotics}, 2014 arXiv:1408.5477.

\bibitem{espositoldb} Van den Broeck C. and Esposito M., {\it Three faces of the second law. I. Master equation formulation},  2010  Phys. Rev. E \textbf{82}, 011143.

\bibitem{cuetara} Bulnes-Cuetara G., Esposito M. and Imparato A., {\it Exact fluctuation theorem without ensemble quantities}, 2014 Phys. Rev. E {\bf 89}, 052119.

\bibitem{agmt} Andrieux D., Gaspard P., Monnai T. and Tasaki S., 2009 {\it The fluctuation theorem for currents in open quantum systems} New J. Phys. {\bf 11}, 043014.

\bibitem{park} Kim K., Kwon C. and Park H., {\it Heat fluctuations and initial ensembles}, 2014 arXiv:1406.7084.

\bibitem{fogedby} Fogedby  H. C. and Imparato A., {\it Heat fluctuations and fluctuation theorems in the case of multiple reservoirs}, 2014 arXiv:1408.0537.

\bibitem{campisi2} Campisi M., H\"anggi P. and Talkner P., {\it Colloquium: Quantum fluctuation relations: Foundations and applications.}, 2011 Rev. Mod. Phys. {\bf 83}, 771.

\bibitem{polettini2} Polettini M., {\it Cycle/cocycle oblique projections on oriented graphs}, 2014 arXiv:1405.0899.

\bibitem{seifert} Seifert U., \textit{Stochastic thermodynamics: principles and perspectives}, 2008 Eur. Phys. J. B {\bf 64}, 423.

\bibitem{polespo} Polettini M. and Esposito M., {\it Irreversible thermodynamics of open chemical networks I: Emergent cycles and broken conservation laws}, 2014 J. Chem. Phys. {\bf 141}, 024117.

\bibitem{kurtz} Ethier S. N. and Kurtz T. G., 1986 {\it Markov processes: Characterization and convergence} (New York: John Wiley \& Sons).

\bibitem{nakanishi} Nakanishi N., 1971 {\it Graph Theory and Feynman Integrals}, (New York: Gordon and Breach).

\bibitem{jaynes} Jaynes E. T., {\it Information theory and statistical mechanics}, 1957 Phys. Rev. {\bf 106}, 620.

\bibitem{prior} Polettini M., {\it Of dice and men. Subjective priors, gauge invariance, and nonequilibrium thermodynamics}, 2013 Proceedings of the 12th Joint European Thermodynamics Conference.
 
\bibitem{gauge} Polettini M., {\it Nonequilibrium thermodynamics as a gauge theory}, 2012
Eur. Phys. Lett. {\bf 97}, 30003.

\bibitem{deficiency} Feinberg  M., {\it Chemical reaction network structure and the stability of complex isothermal reactors-I. The deficiency zero and deficiency one theorems}, 1987  Chem. Eng. Sci. {\bf 42}, 2229.

\bibitem{gregthesis} G. Bulnes-Cuetara, {\it Fluctuation theorem for quantum electron transport in mesoscopic circuits}, arXiv:1310.0620 (2013).

\bibitem{andrieux3} D. Andrieux and P. Gaspard, {\it Temporal disorder and fluctuation theorem in chemical reactions}, Phys. Rev. E {\bf 
77}, 031137 (2008).

\end{thebibliography}
\end{document}